\documentclass[manuscript]{aastex}
\usepackage{emulateapj5}

\usepackage{epsfig}
\usepackage{amsmath}
\usepackage{amssymb}
\usepackage{natbib}
\usepackage{graphicx}

\shorttitle{Decline of kHz QPO amplitude}
\shortauthors{Mukherjee and Bhattacharyya}
\begin{document}

\title{Fractional amplitude of kilohertz quasi-periodic oscillation from 4U 1728--34: 
evidence of decline at higher energies} 

\author{Arunava Mukherjee\altaffilmark{1, 2} and Sudip Bhattacharyya\altaffilmark{1}}
\altaffiltext{1}{Tata Institute of Fundamental Research,
    Mumbai-400005, India; arunava@tifr.res.in}
\altaffiltext{2}{Inter-University Centre for Astronomy and Astrophysics,
    Pune, India; arunava@iucaa.ernet.in}

\begin{abstract}
A kilohertz quasi-periodic oscillation (kHz QPO) is an observationally robust high-frequency timing feature detected from neutron star low-mass X-ray binaries (LMXBs). This feature can be very useful to probe the superdense core matter of neutron stars, and the strong gravity regime. However, although many models exist in the literature, the physical origin of kHz QPO is not known, and hence this feature cannot be used as a tool yet. The energy dependence of kHz QPO fractional rms amplitude is an important piece of the jigsaw puzzle to understand the physical origin of this timing feature. It is known that the fractional rms amplitude increases with energy at lower energies. At higher energies, the amplitude is usually believed to saturate, although this is not established. We combine tens of lower kHz QPOs from a neutron star LMXB 4U 1728--34 in order to improve the signal-to-noise-ratio. Consequently, we, for the first time to the best of our knowledge, find a significant and systematic decrease of the fractional rms amplitude with energy at higher photon energies. Assuming an energy spectrum model, blackbody+powerlaw, we explore if the sinusoidal variation of a single spectral parameter can reproduce the above mentioned fractional rms amplitude behavior. Our analysis suggests that the oscillation of any single blackbody parameter is favored over the oscillation of any single powerlaw parameter, in order to explain the measured amplitude behavior. We also find that the quality factor of a lower kHz QPO does not plausibly depend on photon energy.
\end{abstract}


\section{Introduction}\label{Introduction}

Kilohertz quasi-periodic oscillations (kHz QPOs), discovered in 1996 with {\it Rossi X-ray Timing 
Explorer} ({\it RXTE}; \citet {vanderKlisetal1996, Strohmayeretal1996}), are the fastest variability 
features known till date in low-mass X-ray binaries (LMXBs). They have been observed from many neutron star
LMXBs, although for a given source they are not always detected \citep {Bhattacharyya2010, vanderKlis2006}. These QPOs often occur in pairs, and the twin peaks usually move together in the frequency range of 
$\sim 400-1200$ Hz. The higher frequency QPO is known as the upper kHz QPO (frequency $\nu_{u}$), and the lower frequency QPO is called the lower kHz QPO (frequency $\nu_{l}$). The frequency differences of these QPOs 
($\Delta \nu \equiv \nu_{u} - \nu_{l}$) tend to cluster around the neutron star spin frequency or half of it (\citet {vanderKlis2006} and references therein; but see \citet{MendezBelloni2007}). 
The high frequencies of kHz QPOs point toward the dynamical timescale within a few Schwarzschild
radii of the neutron star \citep{Barretetal2005a, Barretetal2005b, Barretetal2006, 
vanderKlis2006, Bhattacharyya2010}. Therefore, this observationally robust timing feature
can be useful (1) to measure the neutron star parameters, which, in turn, is essential
to understand the nature of super-dense degenerate matter (e.g., \citet{Bhattacharyya2010}
and references therein; \citet{Petri2011}); and 
(2) to probe the strong gravitational field regime \citep{Psaltis2008}.
However, although many models are available in the literature, the physical origin of 
kHz QPOs is still not known (e.g., \citet{Linetal2011, Torok2009, vanderKlis2006}),
and hence we cannot yet use this promising feature as a tool.
Many proposed models for this timing feature have attempted to explain the frequencies
\citep{Milleretal1998, StellaVietri1998, StellaVietri1999, LambMiller2003, 2001AcPPB..32.3605K, 
AbramowiczKluzniak2001, Wijnandsetal2003, Leeetal2004, Zhang2004, Zhangetal2006, 
Mukhopadhyay2009, AlparandPsaltis2008, Stuchliketal2011}.
Some of these models involve several general relativistic frequencies at preferred radii and the 
neutron star spin frequency; in some cases beating and/or resonances among these frequencies.
Other models include the association of kHz QPOs with 
accretion through a non-axisymmetric magnetic boundary layer
in the unstable regime \citep{Romanovakulkarni2009};
attempts to connect the kHz QPOs with the trapped two-armed
nearly vertical oscillations in vertically isothermal disks
with toroidal magnetic fields \citep{Kato2011}; attribution of the upper kHz QPO
profile to the radial extent of the kHz QPO emission region associated with the
transitional layer at the magnetosphere-disk boundary \citep{Wangetal2011}, etc.

However, modeling only the frequencies gives an incomplete picture; probing radiative transfer
and further modulation processes is essential to constrain the existing models for
understanding the physical mechanism giving rise to kHz QPOs. For example, the models based
on frequencies tentatively suggest some locations of kHz QPO origin. These locations
could be at certain radii of the accretion disk, such as the innermost-stable-circular-orbit 
radius, sonic point radius, etc. (see \citet{vanderKlis2006} and references therein).
However, one needs to independently verify these proposed locations.
Since different energy spectral components originate from different locations, 
such as disk, boundary layer, corona, etc., 
a connection found between a kHz QPO property and a spectral component could
provide this independent verification.
Therefore, it is essential to study the energy dependence of kHz QPO properties.

In this paper, we study the energy dependence of fractional rms amplitude of kHz QPOs.
This property is a measure of QPO strength, and hence studying energy dependent kHz QPO
property  can be useful to probe which spectral component primarily contributes
to this timing feature. This spectral connection of kHz QPO has been discussed by
some authors. For example, \citet{Mendez2006} suggested that,
while the kHz QPO frequencies are plausibly determined at the disk,
this feature is modulated at the high-energy spectral component (e.g., corona, 
boundary layer, etc.). It has been reported by several authors that the kHz QPO fractional 
rms amplitude increases with photon energy at lower energies, and then plausibly 
saturates (\citet{vanderKlis2006} and references therein; \citet{Gilfanovetal2003}). 
Several authors have theoretically computed the energy dependence of rms amplitude of
Comptonizing component variability (e.g., \citet{Cabanacetal2010, 2005MNRAS.363.1349G, 2005MNRAS.360..825Z, LeeMiller1998}).
For example, \citet{Cabanacetal2010} have demonstrated that oscillating hot thermal corona 
may give rise to an overall increase in rms variability with photon energy.
However, the above mentioned saturation at higher energies is not observationally established for many sources
due to the lack of sufficient signal-to-noise ratios (S/Ns) at higher energies. For example,
while the higher energy data points of Fig. 10 of \citet{Gilfanovetal2003} are consistent with 
a flat fractional rms amplitude versus energy behavior, the errors of these data points are quite high.

With this motivation, we try to measure the fractional rms amplitude versus energy behavior
of lower kHz QPOs with improved S/N. We combine tens of lower kHz QPOs from the neutron star
LMXB 4U 1728--34, in order to improve the S/N (e.g., \citet{FordvanderKlis1998, MendezvanderKlis1999, Disalvoetal2001}). We choose lower kHz QPO, because this narrow QPO is more frequent and easier to detect than the relatively broad and weak upper kHz QPO. We find that the fractional rms amplitude systematically decreases with energy at higher energies. In order to understand this new finding, we compare the data with models involving a blackbody+powerlaw energy spectrum. This comparison suggests
that the observed behavior of fractional rms amplitude may be reproduced with a fluctuation
of the blackbody spectral component. In addition, we find that the quality factor (Q)
of lower kHz QPO does not plausibly depend on energy.

In \S~\ref{DataAnalysis} and \S~\ref{Results}, we describe our data analysis technique in detail
and display our results, respectively. In \S~\ref{Models}, we describe the models and discuss
which ones are favored. Finally, in \S~\ref{Discussion}, we summarize our results and give implications.

\section{Data Analysis}\label{DataAnalysis}

In order to study the energy dependence of fractional rms amplitude
associated with lower kHz QPOs, we retrieved all {\it RXTE}
proportional counter array (PCA) pointed observational
data of the source 4U 1728--34 for the period of April 13, 2000 to the end
of December, 2009 ($\approx 272$ ks of cleaned exposure time) corresponding
to the PCA gain epoch 5. Note that gain, and hence the energy-channel
conversion\footnote{http://heasarc.gsfc.nasa.gov/docs/xte/e-c\_table.html},
has changed during the entire {\it RXTE} lifetime. Therefore, we restrict
our analysis to the data of a single epoch in order to make the energy dependence
measurement more reliable. Although even within an epoch the energy-channel conversion evolves,
such changes would be negligibly small to significantly affect the final conclusion
of the analysis. We choose epoch 5, because it has the largest amount of data among
all the epochs. We consider only the science event files with time resolution
$\sim 122 \mu$s and having continuous exposure time of at least 400 s for searching kHz QPOs.
We do not apply any filtering based on energies, PCUs, Xenon layers, etc.
However, following the usual practice, we filter the data based on time
in order to remove thermonuclear X-ray bursts, data gaps, and observed intensity
increase/decrease due to instrumental effects (especially due to start or stop of a PCU).

It has been observed that all the kHz QPOs (which are strong and narrow)
are confined to a small portion, i.e., lower banana, of the color-color diagram (CD) of
4U 1728--34 \citep{Disalvoetal2001}. This suggests that they are of the same spectral origin.
The strong inter-dependence found between Q-values and frequencies \citep{Barretetal2006}
of lower kHz QPOs from 4U 1728--34 suggests that all the QPOs considered by us are of
same physical origin. These justify the combination of many kHz QPOs 
(as mentioned in \S~\ref{Introduction}).

We follow a few steps to probe the energy dependence of lower kHz QPO amplitudes. 
In the first step, we collect all the significant kHz QPOs
from the event files (with at least 400 s of data) using a blind search. In order to
do this, we compute a Leahy-normalized \citep{Leahyetal1983} power density spectrum (PDS)
using discrete Fourier transform (DFT)\footnote{DFT follows the same statistics of
Leahy-normalization \citep{vanderKlis1989}.} from each 10 s of data for a given 
event file \citep{vanderKlis1989}.
The Nyquist frequency and frequency resolution of each such PDS (from 10 s of data) 
are $2048$ Hz and $0.1$ Hz, respectively \citep{vanderKlis1989}. Then all (say, $N$)
the PDSs from a given event file are averaged, and we search for kHz QPOs in the frequency 
range of $400-1400$ Hz in this mean PDS. In order to search 
effectively, several (say, $W$) adjacent frequency bins are combined.
The noise powers in this PDS follow a $\chi^{2}$ distribution with $2NW$ degrees of freedom
\citep{vanderKlis1989}. For a putative peak, we compute the single trial
probability ($q$) of occurrence for a power greater than or equal to the peak power
by chance from the noise power distribution. Note that we consider the peak power, and do
not fit the peak with a function (say, Lorentzian) for significance calculation.
The probability $q$ is multiplied with the number of trials ($N_{\rm trial}$) in order to
obtain the final probability $\epsilon$ ($= q N_{\rm trial}$). We use $N_{\rm trial}
= 20480$, i.e., the original total number of frequency bins in a mean PDS.
We consider a putative peak as a kHz QPO, if $\epsilon \leq4.65 \times 10^{-4}$
(i.e., at least $3.5 \sigma$ significant).
We detect tens of kHz QPOs from 4U 1728--34 using the above procedure.

In the second step, described in this paragraph, 
we identify the lower kHz QPOs from the QPOs detected
using the above mentioned criterion.
If twin kHz QPO peaks appear in the mean PDS from an event file, the
identification of the lower kHz QPO requires no further effort. If there is only
one peak for an event file, then we consider it as a lower kHz QPO, if its
$Q \ge 40$. This is because,  the Q vs frequency diagram of 4U 1728--34
\citep{Barretetal2006} clearly shows that only the lower kHz QPOs can have Q-values
greater than 40, and all the upper kHz QPOs have smaller Q-values. In order to
estimate the Q-values of kHz QPOs, we first use the shift-and-add technique
(e.g., \citet{Mendezetal1998,Barretetal2006}) within each event file considering
the entire PCA energy range. This technique mostly corrects for the
centroid frequency drift. Note that such uncorrected drift makes the measured
Q-value smaller than the actual value. After application of shift-and-add, we fit each PDS
containing a detected kHz QPO with a `constant + powerlaw + Lorentzian' model;
constant takes care of the Poissonian white noise, powerlaw takes care of the low
frequency red noise, and a Lorentzian describes the kHz QPO. We obtain the best-fit
centroid frequency ($\nu$) and the best-fit full width at half maximum (FWHM)
from the Lorentzian component, and compute the Q-value ($\nu/{\rm FWHM}$).
Then we identify the lower kHz QPOs using the criterion $Q \ge 40$.
From the entire $\approx 272$ ks of data, 40 lower kHz QPOs (each from one
event file) are identified
spanning $86.32$ ks of exposure time, and are considered for further analysis.

These 40 lower kHz QPOs are now to be combined in order to study the energy
dependence of fractional rms amplitude with improved S/N. However, before doing this,
we perform a few preliminary analyses in the third step, described in this paragraph.
In this step, we still use the entire PCA energy range. We consider non-overlapping
400-second segments from each of 40 event files with a lower kHz QPO.
Now we collect only those 400-second segments, in each of which a lower kHz QPO peak
with $\epsilon \leq 2.7 \times 10^{-3}$ (corresponding to $3\sigma$) exists.
To calculate the $\epsilon$ for a given 400-second segment, we further divide the segment
into 40 segments of 10-second intervals, calculate PDS for each 10-second interval, average
these 40 PDSs, and use $N=40$ and $N_{\rm trial} = 10000/W$. Note that this procedure, as well
as the definitions of $\epsilon$, $N$, $W$ and $N_{\rm trial}$ are given in
the third paragraph of the current section. We find 173 number of 400-second segments
with strong lower kHz QPO peaks using the above mentioned criterion on
$\epsilon$. The mean of these peaks is $\nu_{\rm mean} = 807.8$ Hz.
We also calculate $\nu_{\rm diff}$, which is the separation between $\nu_{\rm mean}$ and
the centroid frequency of an individual lower kHz QPO, for each 400-second segment.
We use these 173 segments, $\nu_{\rm mean}$, and $\nu_{\rm diff}$ values for further analysis.

In the fourth step, described in this and the next three paragraphs, 
we compute lower kHz QPO fractional rms amplitudes for several
chosen energy ranges after combining the data of 173 number of 400-second segments
(see the previous paragraph). We consider a set of PCA absolute channel ranges
5--8, 9--11, 12--13, 14--15, 16--17, 18--21, 20--25, 22-25, 22--29, 24--31, 26--33
and 26--49 corresponding to the energy ranges 2.06--3.68, 3.68--4.90, 4.90--5.71,
5.71--6.53, 6.53--7.35, 7.35--8.98, 8.17--10.63, 8.98--10.63, 8.98--12.28, 9.81--13.11,
10.63--13.93 and 10.63--20.62 keV,
respectively\footnote{http://heasarc.gsfc.nasa.gov/docs/xte/e-c\_table.html},
in order to study the energy
dependence of fractional rms amplitude. We do not use proportional counter unit 0 (PCU0) in
this step, because the energy-channel conversion for PCU0 is somewhat different
from that of the other four PCUs during epoch 5. For each energy range,
we compute a mean Leahy-normalized PDS (in the same way as described earlier in this
section) for each of 173 number of 400-second segments. Then, we shift each of 173
lower kHz QPO peaks by $\nu_{\rm diff}$ (the value for the entire PCA energy range; 
see the previous paragraph) to align them at $\nu_{\rm mean}$, and add them together
to obtain a grand average Leahy-normalized PDS for every energy range. 

Now the question is whether the centroid frequencies of the kHz QPOs change with energy.
Such a change might affect our grand average Leahy-normalized PDS and further results.
We cannot check it directly for individual kHz QPOs, because the statistics is not
often adequate to detect an individual kHz QPO in a small energy range. So we check it
in the following way. If we consider that centroid frequency of each of the individual kHz QPOs
is energy dependent, then there are two extreme possibilities.
(1) The first possibility: the centroid frequencies of all these kHz QPOs either
increase or decrease at higher energy in a similar way. In this case, the
$\nu_{\rm mean}$ value is also expected to either increase or decrease at higher energy,
since we do not recalculate $\nu_{\rm diff}$ separately for each energy range.
But we find that the $\nu_{\rm mean}$ value remains the same in all the energy bands
(Figs.~\ref{PDS_SAT1} and \ref{PDS_SAT2}).
(2) The second possibility: the centroid frequencies of some kHz QPOs increase, and
those of some other decrease with the increase of energy, in such a way
that all the $\nu_{\rm mean}$ values at various energy ranges remain the same.
In this case, although the centroid frequency would not change, the width
of the merged kHz QPOs (i.e., after applying the shift-and-add method) is expected to
systematically increase resulting in a lower and lower Q-value, at higher and
higher energies. But we find that all the measured Q-values are quite close to 
each other (within the errors) and no systematic variation is observed.
The fact that both $\nu_{\rm mean}$ values and Q-values remain quite similar in all the
energy ranges (Figs.~\ref{PDS_SAT1}, \ref{PDS_SAT2} and \ref{RMS-Qfactor-vs-Energy}) 
suggests that the centroid frequencies of the kHz QPOs do not change with energy.

The shift-and-add technique to obtain a grand average Leahy-normalized PDS (mentioned above)
is a standard method to improve the S/N of kHz QPOs and
to correct their Q-value (e.g., \citet{Barretetal2006, Mendez2006, Mendezetal2001,
MendezvanderKlis1999, Mendezetal1998, MukherjeeBhattacharyya2011b}). The grand average
Leahy-normalized PDS for each energy range is then fitted with a `constant + powerlaw +
Lorentzian' model (as before), and the fractional rms amplitude and the Q-value are computed from
the best-fit Lorentzian parameters (see the equation 4.10 of \citet{vanderKlis1989} for a general
rms amplitude formula).

In the fifth step, we correct the above mentioned fractional rms amplitudes
for the background levels. First, we compute the background
count rate in each of the chosen energy ranges using the appropriate channel range
of the corresponding standard-2 data files. In order to do this, we use the
PCA background model file for bright sources, and the `FTOOLS' command `pcabackest'.
Then, we compute the background corrected fractional rms amplitudes ($R_{c}$) from the
uncorrected fractional rms amplitudes ($R_{uc}$), the total count rates ($I$) and
the background count rates ($B$) using the formula
$R_{c} = R_{uc} \times I/(I-B)$ \citep{Munoetal2002}.

Apart from the energy dependence of fractional rms amplitude, we estimate
how the Q-value of the lower kHz QPO depends on energy. This is computed from
the above mentioned fitting of the grand average Leahy-normalized PDS for
each energy range. The ratio of the best-fit centroid frequency to the
best-fit FWHM of the Lorentzian model component gives the Q-value.
Here we note that while the shift-and-add technique corrects for 
a large error in Q-value due to the
centroid frequency drift, this technique also introduces a small
error in Q-value, because the centroid frequency is shifted without
changing the FWHM.

\section{Results}\label{Results}

The frequencies of the 173 detected lower kHz QPOs, which are used to 
study the energy-dependence of fractional rms amplitude (\S~\ref{DataAnalysis}), span 
the range of $669.4-912.5$ Hz (mean ($\nu_{\rm mean}$) = $807.8$ Hz; 
median = $810.2$ Hz, and standard deviation = $57.8$ Hz).
In Figs.~\ref{PDS_SAT1} and \ref{PDS_SAT2}, the grand average Leahy-normalized PDSs
(see \S~\ref{DataAnalysis}) for 12 energy ranges are shown in 12 panels. Each panel is
for the same exposure (69.2 ks), and in each panel the total count rate and the background
count rate are mentioned. These figures indicate that, while there are visibly prominent
kHz QPO peaks for the lower energy bands, such a peak is not visible for the highest
energy band. 

In Fig.~\ref{RMS-Qfactor-vs-Energy}, we plot background corrected fractional rms amplitude
versus energy. The amplitude 
increases with energy at lower energies, and systematically decreases 
(seen from the overlapping energy ranges) above $\approx 10$ keV. 
Note that, if for an energy range the
lower kHz QPO peak is not clearly visible (see Fig.~\ref{PDS_SAT2}; panels with blue model curves), 
we compute an upper limit
to the fractional rms amplitude at $\nu_{\rm mean}$ (see \S~\ref{DataAnalysis}) instead of
estimating the amplitude from the best-fit Lorentzian. Our estimated 
$1\sigma$ and $3\sigma$ upper limits of fractional rms amplitude in the energy range
10.63--20.62 keV are 2.37\% and 3.66\%, respectively.
If we compare the fractional rms amplitude value ($ = 14.38\% \pm 1.71$\%) in the 7.35--8.98 keV range
with the $3\sigma$ upper limit ($ = 3.66$\%) in the 10.63--20.62 keV range, we find
a $6.3\sigma$ significant drop in the amplitude value. This quantifies the decrease of
the lower kHz QPO fractional rms amplitude at higher energies (see Fig.~\ref{RMS-Qfactor-vs-Energy}).

The initial increasing part of the fractional rms amplitude is fitted with a linear
and a powerlaw model, the best-fit curves of which are very similar to each other
(see Fig.~\ref{fit_RMS-Qfactor-vs-Energy_extpl}). This figure, with extrapolated model curves, 
shows that the increasing trend of lower kHz QPO amplitude at lower energies cannot plausibly
explain the upper limit in the 10.63--20.62 keV range. For example, the extrapolated fractional 
rms amplitude value at 10.63--20.62 keV from the linear model is $30.08\pm6.80$, which implies
a $3.89\sigma$ significant higher value compared to the $3\sigma$ upper limit of the observed
fractional rms amplitude.

While the fractional rms amplitude changes with energy, Q-value does not 
(Fig.~\ref{RMS-Qfactor-vs-Energy}, lower panel). 
As mentioned in \S~\ref{DataAnalysis}, while the shift-and-add technique
corrects for a usually large error in Q-value due to the evolution of the centroid
frequency, this technique introduces a small error. This is because, while a kHz QPO
peak is shifted to a new frequency ($\nu_{\rm mean} = 807.8$ Hz), the FWHM of this peak
remains the same. Given that the standard deviation of the centroid frequency distribution
of our detected lower kHz QPOs is 57.8 Hz and $\nu_{\rm mean} = 807.8$ Hz, this error can
be estimated to be $\sim 7$\% of the typical value of Q $\sim 80$. Therefore,
this error is small compared to the $1\sigma$ error (which is typically $\sim 25\% - 45\%$) 
from Lorentzian fitting of the QPO peaks. Therefore, the error due to frequency shift
does not change our conclusion regarding the energy independence of Q-values, 
and hence we do not make an attempt to correct it.

\section{Comparison with models}\label{Models}

This paper shows a prominent and systematic decrease of the 
lower kHz QPO fractional rms amplitude at higher energies,
for the first time to the best of our knowledge. We, therefore, try to find out
if such a decrease is at all theoretical expected. For example, could certain parameter
values of the usual energy spectra of neutron star LMXBs explain it qualitatively?
Before testing this, let us briefly mention the X-ray components of these
sources, as we understand currently. Neutron star LMXBs are believed to have two 
primary X-ray emitting regions, an accretion disk and a boundary
layer. Both these regions are expected to be optically thick, and therefore to emit
blackbody radiation. Furthermore, one (or more) of these components may be 
covered with a corona (coronae). Such a corona may reprocess (Comptonize) the
blackbody photons. The amount of reprocessing depends on the optical depth of
the corona, as well as the extent of coverage (e.g., full versus partial).
As a result, the observed spectrum either from the disk and/or the boundary layer
can be a blackbody, or Comptonized, or a sum of both. Observationally, however,
no spectral model usually uniquely describes the data. Moreover, it is not usually 
clear, where various components of a given model originate from. Therefore,
although many works over a few decades (e.g., \citet{Mitsudaetal1989, MitsudaandDotani1989,
Whiteetal1988, ChurchBalucinskaChurch2001, ChristianSwank1997, MaccaroneCoppi2003,
MaitraBailyn2004, Gilfanovetal2003, Oliveetal2003, Wijnands2001, Barret2001,
Linetal2007, Linetal2010, MukherjeeBhattacharyya2011a}) 
support the general understanding mentioned above, the details are still unknown.

In order to have a basic understanding (i.e., as much as possible independent of detailed models)
of the lower kHz QPO fractional rms amplitude versus energy behavior found by us, we choose
the simplest two-component spectral model, which is based on the above mentioned
description of X-ray components. This is a blackbody+powerlaw model, in which powerlaw
usually represents the Comptonized component. \citet{Taranaetal2011} and \citet{SeifinaTitarchuk2011}
have successfully fitted the continuum spectra from 4U 1728--34 with a blackbody+Comptonization model.
In order to check whether a blackbody+powerlaw works, we fit the continuum 
spectra of a few data files (having prominent kHz QPOs) with this model, as well as
with a blackbody+Comptonization ({\tt bbodyrad+compTT} in XSPEC) model. We find
that, not only both the models give acceptable fits, but the powerlaw component
represents the Comptonization component reasonably well, as verified from the
respective fluxes and spectral component curves. This indicates that a blackbody+powerlaw
model is appropriate for the 4U 1728--34 spectra.

Before describing our model, here we mention and discuss the assumptions involved in our
modeling and interpretation, some of them recapitulated from the data analysis side and
some other extra assumptions from the model side.
(1) We assume that the spread in energy spectra (e.g., soft colors and
hard colors in a color-color diagram (CD); \citet{vanderKlis2006}) 
and in intensity from our data files containing lower kHz QPOs do not
significantly affect the robustness of the energy behavior of fractional rms
amplitude reported by us (Fig.~\ref{RMS-Qfactor-vs-Energy}). This is justified because
of the following reasons. (a) All the QPOs (which are strong and narrow)
considered by us are confined to a small portion, i.e., lower banana,
of the CD of 4U 1728--34 \citep{Disalvoetal2001}, as mentioned in \S~\ref{DataAnalysis}.
(b) Given the small fractional rms amplitudes ($\sim 5-14$\%) of lower kHz QPOs in the 2-16 keV
energy range, which is used to compute the CD and HID, the entire X-ray energy spectrum 
does not contribute to these QPOs. Hence a small spread in CD does not necessarily indicate
different physical origin of different lower kHz QPOs.
(c) The strong inter-dependence found between Q-values and frequencies of lower kHz QPOs
from 4U 1728--34 (along with several other atoll sources; \citet{Barretetal2006})
suggests that all the QPOs considered by us are of same physical origin.
These support the robustness of our reported rms-energy behavior.
(2) We assume that the energy dependence of lower kHz QPOs does not change with frequency.
This may be a reasonable assumption, 
because these QPOs with different frequencies are likely
to have originated from the same physical process.
(3) Our model considers the sinusoidal fluctuation of a single parameter of the above mentioned
blackbody+powerlaw model. 
The sinusoidal fluctuation is somewhat justified, because according to some models,
the original oscillations are sinusoidal, and the broadening happens because of
a decoherence mechanism (say, damping; \citet{vanderKlis2006} and references therein).
Here we note that a sinusoidal signal was also assumed by \citet{LeeMiller1998}.
We further note that, even if the intrinsic oscillations of lower kHz QPOs are a sum
of many sinusoids originated separately, our modeling will still be useful to connect a spectral component
to these QPOs.
(4) Our model does not involve any response matrix. Although application of
a response matrix may slightly change the model rms-energy curves
quantitatively, the qualitative nature of these curves should not change.

Now we describe our model. As previously mentioned, the total time averaged flux
(say, $<S(E,t)>$; $E$: photon energy, $t$: time) of the source can be fitted well
with a combination of blackbody ($S_{\rm BB}$) and powerlaw ($S_{\rm PL}$).
Each of blackbody and powerlaw has two parameters: normalization ($N_{\rm BB}$) 
and temperature ($T_{\rm BB}$) for the former, and normalization ($N_{\rm PL}$) 
and photon index ($\alpha_{\rm PL}$) for the latter. We try to 
investigate if sinusoidal fluctuation in one of these parameters 
(while keeping the other three parameters non-fluctuating) can reproduce the
observed fractional rms amplitude versus energy behavior. Here as an example,
we consider the case of fluctuating blackbody temperature.
\begin{eqnarray*}
N_{\rm H} \times S(E,t) &=& N_{\rm H} \times [S_{\rm BB} (E, N_{\rm BB}, T_{\rm BB}(t)) \\
 &+& S_{\rm PL} (E, N_{\rm PL}, \alpha_{\rm PL})] \\
\text{where}, T_{\rm BB}(t) &=& T_{\rm BB0} + T_{\rm BB1} Sin(\omega t) \\
N_{\rm H} \times <S(E,t)> &=& N_{\rm H} \times [<S_{\rm BB} (E, N_{\rm BB}, (T_{\rm BB0} \\
 &+& T_{\rm BB1} Sin(\omega t)))> + S_{\rm PL} (E, N_{\rm PL}, \alpha_{\rm PL})] \\
N_{\rm H} \times \sigma_{S} (= \sigma_{NS}) &=& N_{\rm H} \times \surd (<S(E, t)^{2}> - <S(E, t)>^{2}) \\
\text{fracRMS} &=& \frac{\sigma_{NS}}{N_{\rm H} \times <S(E,t)>} \\
 &=& \frac{\surd (<S(E, t)^{2}> - <S(E, t)>^{2})}{<S(E,t)>} \\
\end{eqnarray*}
which is independent of $N_{\rm H}$ (neutral hydrogen column density). Here, fracRMS is our model
fractional rms amplitude. Similarly, we compute the model fractional rms amplitude in each case
when any one of the other three parameters varies sinusoidally in time.

We note that the blackbody flux and the powerlaw flux are nonlinear functions of $T_{\rm BB}$
and $\alpha_{\rm PL}$, respectively. Therefore, if $T_{\rm BB}$ (or $\alpha_{\rm PL}$) fluctuates sinusoidally
(as we assume), the variation of the blackbody flux (or powerlaw flux) will not be strictly
sinusoidal, and will contain higher harmonics. Here we show that this flux variation is
sinusoidal (which we assume in our modeling) when $T_{\rm BB}$ (or $\alpha_{\rm PL}$) fluctuation is small
(e.g., $T_{\rm BB1} \ll T_{\rm BB0}$, for the varying blackbody temperature), and is therefore considered up to the first order:
\begin{eqnarray*}
<S(E,t)> &=& <S_{\rm BB} (E, N_{\rm BB}, (T_{\rm BB0} + T_{\rm BB1} Sin(\omega t)))> \\
&+& <S_{\rm PL} (E , N_{\rm PL}, \alpha_{\rm PL})> \\
&=& <S_{\rm BB} (E, N_{\rm BB}, T_{\rm BB0})> \\
&+& <S_{\rm BB1} (E, N_{\rm BB}, T_{\rm BB0}, {\rm T_{BB1}}) Sin(\omega t)> \\
 &+& O[\frac{T_{\rm BB1}}{T_{\rm BB0}}]^2 + S_{\rm PL} (E, N_{\rm PL}, \alpha_{\rm PL}) \\
&=& S_{\rm BB} (E, N_{\rm BB}, T_{\rm BB0}) + S_{\rm PL} (E, N_{\rm PL}, \alpha_{\rm PL})\\
& & \text{(which is correct up to first order in $[\frac{T_{\rm BB1}}{T_{\rm BB0}}]$)} \\
\end{eqnarray*}
The last step is justified because the observations are several orders of magnitudes longer
than the oscillation period $T = 2\pi /\omega$. This last step shows that, $<S(E,t)>$ $= S(E)$; 
which is simply the sum of {\tt XSPEC} model components ({\tt bbodyrad+powerlaw}) used to
fit the time averaged spectrum. Finally, note that this first order approximation is
not required when normalization of blackbody or normalization of powerlaw varies.

Considering the above procedure, our model fractional rms amplitude has five parameters: 
$N_{\rm BB}$, $T_{\rm BB}$, $N_{\rm PL}$, $\alpha_{\rm PL}$ and $A$, 
the last one is the fractional peak amplitude of one of the first four parameters,
which is fluctuating (e.g., $\frac{T_{\rm BB1}}{T_{\rm BB0}}$ in the case discussed above for
the varying blackbody temperature). While computing the models, we consider following ranges of our parameters:
$N_{\rm BB} = 0.5-50$, $T_{\rm BB} = 0.3-3.0$ keV, $N_{\rm PL} = 0.5-50$ photons cm$^{-2}$ s$^{-1}$ keV$^{-1}$
(at 1 keV), $\alpha_{\rm PL} = 1.5-3.5$ and $A = 0.001-0.9$. 
We note that the formula for blackbody photon count rate
per cm$^2$ per keV is $(N_{\rm BB} \times 1.0344 \times 10^{-3} \times E^2)/(\exp(E/T_{\rm BB})-1)$,
while that for powerlaw photon count rate per cm$^2$ per keV is
$N_{\rm PL} E^{-\alpha_{\rm PL}}$, with energy $E$ in keV. The best-fit parameter
values from our fitting of a few 4U 1728--34 spectra (see earlier in this section), 
and from \citet{Taranaetal2011} and \citet{SeifinaTitarchuk2011} are within the above ranges.

For a set of five model-parameter values, we numerically compute the fractional rms amplitude, as outlined above, systematically in each of the observed energy ranges for the sinusoidal oscillation in time of each of $N_{\rm BB}$, $T_{\rm BB}$, $N_{\rm PL}$ and $\alpha_{\rm PL}$. For each of these four cases, we produce many `fractional rms amplitude versus energy' model curves for various sets of parameter values, in order to thoroughly explore if some curves for a given case can reasonably describe the data. We show some examples of these model curves in Figs.~\ref{khzqpoBBpl_BBtemp_1.ps}, \ref{khzqpoBBpl_BBNorm_1.ps}, \ref{khzqpoBBpl_plindex_1.ps}
and \ref{khzqpoBBpl_plNorm_1.ps}. In each of these figures, for the variation of a single model parameter, we show the example curves for various sets of parameter values (mentioned in the figure-caption). Therefore, these figures give an idea about the nature of energy-dependence of fractional rms amplitudes for various parameter values for the sinusoidal oscillation in time of a given parameter. The blue dashed curve in each figure appears to be the closest to the data among our computed model curves. These figures show that the sinusoidal oscillation of a blackbody parameter describes the data well, and hence the observed energy-dependence of fractional rms amplitude is expected. We note that testing this was the main aim of modeling reported in this paper, as mentioned in the first paragraph of this section. Figs.~\ref{khzqpoBBpl_BBtemp_1.ps},
\ref{khzqpoBBpl_BBNorm_1.ps}, \ref{khzqpoBBpl_plindex_1.ps} and \ref{khzqpoBBpl_plNorm_1.ps} also suggest that the oscillation of a single blackbody parameter (temperature or normalization) describes the data better than the variation of a single powerlaw parameter (index or normalization). We further note that the small value of $A$ corresponding to the blue dashed curve of Fig.~\ref{khzqpoBBpl_BBtemp_1.ps} satisfies the assumption of small fluctuation of the blackbody temperature, as mentioned earlier in this section.

\section{Discussion}\label{Discussion}

\begin{figure*}
\centering
\begin{tabular}{c}
\hspace{-1.0cm}
\includegraphics*[width=0.85\textwidth]{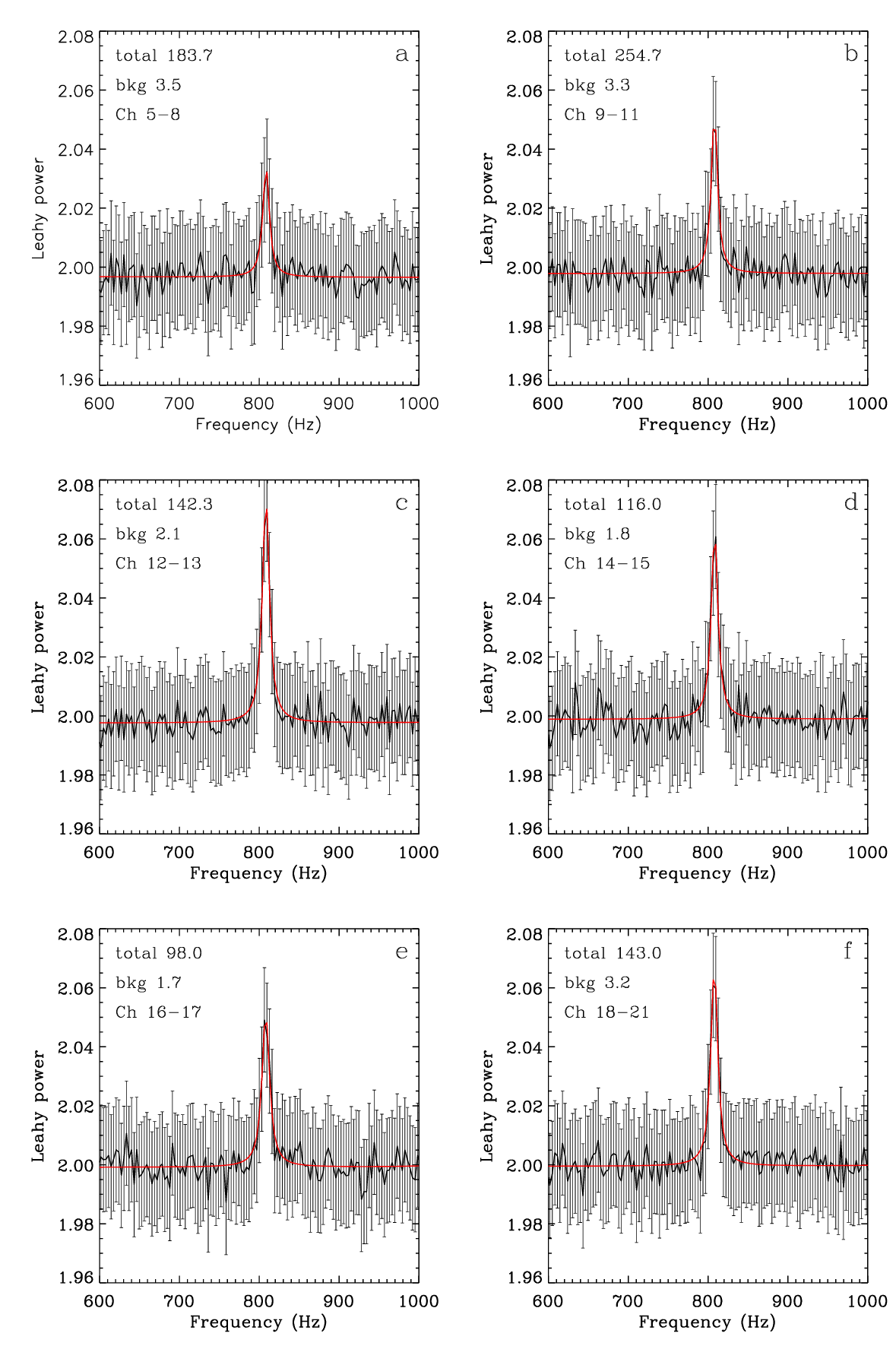}
\end{tabular}
\caption{Leahy-normalized power density spectra (PDS; with $1\sigma$ errors) 
of 4U 1728--34 after using the shift-and-add 
technique for each of the energy ranges 2.06--3.68 keV (panel {\it a}), 3.68--4.90 keV (panel {\it b}), 
4.90--5.71 keV (panel {\it c}), 5.71--6.53 keV (panel {\it d}), 
6.53--7.35 keV (panel {\it e}) and 7.35--8.98 keV (panel {\it f}; see 
\S~\ref{DataAnalysis}). In each panel, the average total count rate (i.e. without background subtraction) 
used to compute the PDS is denoted with `total', the average background count rate is denoted with
`bkg', and the PCA absolute channel range is denoted with `Ch'. Furthermore, the frequency
resolution and exposure in each panel are 3.2 Hz and 69.2 ks, respectively. 
A lower kHz QPO peak is clearly seen in each panel.
The best-fit `constant + powerlaw + Lorentzian' 
model for each energy range is shown with a red curve (\S~\ref{DataAnalysis}). 
The fractional rms amplitude and the Q-value of the lower kHz QPO in each energy range 
are obtained from the Lorentzian model component (\S~\ref{DataAnalysis} and \ref{Results}).
\label{PDS_SAT1}}
\end{figure*}

\begin{figure*}
\centering
\begin{tabular}{c}
\hspace{-1.0cm}
\includegraphics*[width=0.85\textwidth]{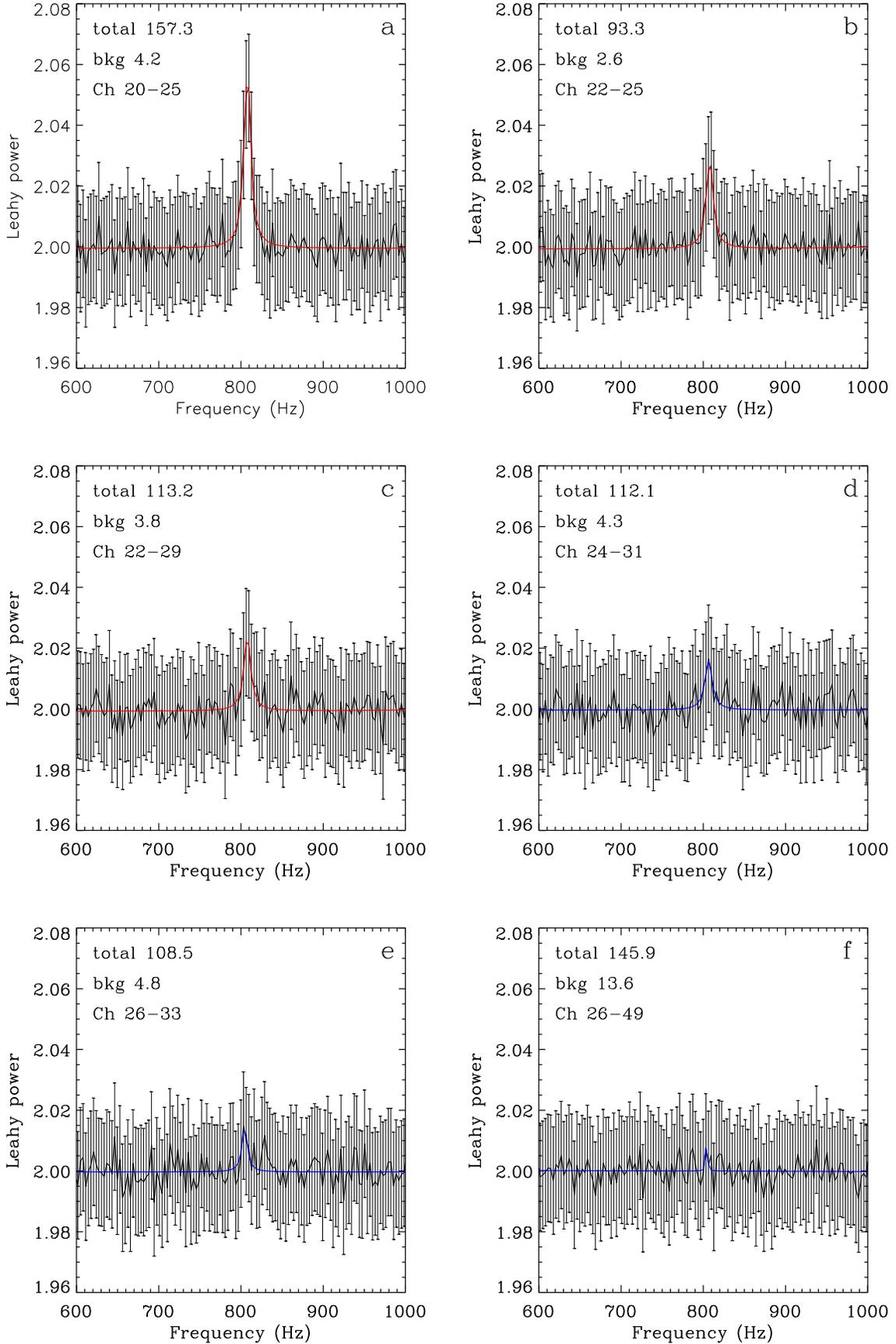}
\end{tabular}
\caption{Same as Fig.~\ref{PDS_SAT1}, but for the energy ranges 8.17--10.63 keV (panel {\it a}), 
8.98--10.63 keV (panel {\it b}), 8.98--12.28 keV (panel {\it c}), 9.81--13.11 keV (panel {\it d}), 
10.63--13.93 keV (panel {\it e}) and 10.63--20.62 keV(panel {\it f}; see \S~\ref{DataAnalysis}). 
In the panels {\it d}, {\it e} and {\it f}, the lower kHz QPO peak is not clearly seen, and hence
the best-fit model curves in these panels are marked with a different color (blue). For each of these
three energy ranges, an upper limit to the lower kHz QPO fractional rms amplitude is computed
(see \S~\ref{Results}).
\label{PDS_SAT2}}
\end{figure*}

\begin{figure*}
\centering
\includegraphics*[width=0.8\textwidth]{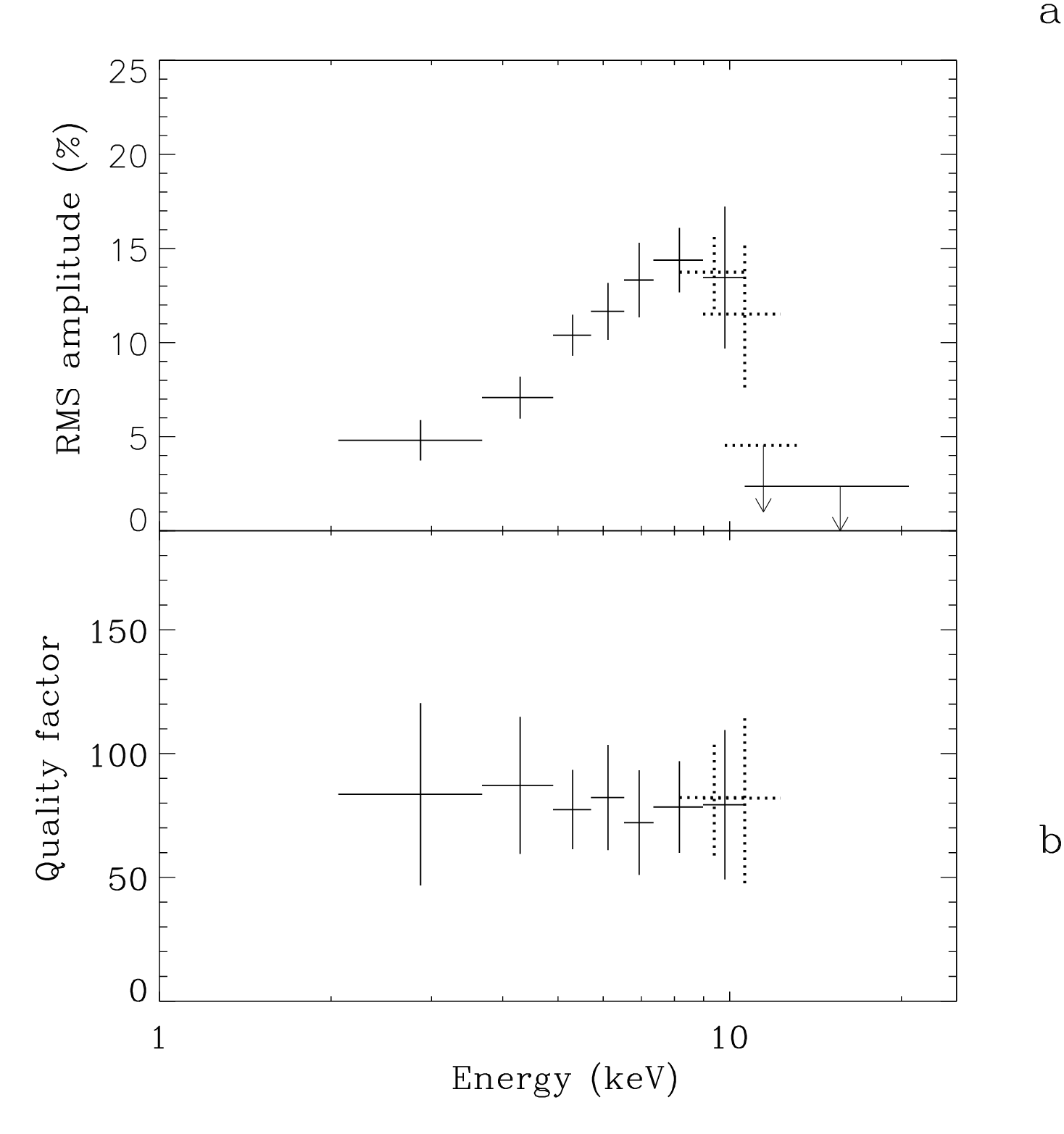}
\caption{Energy dependence of the background corrected fractional rms amplitude and the 
Q-value of the lower kHz QPO from 4U 1728--34. In each panel, solid horizontal lines
denote a set of energy ranges which are mutually non-overlapping, and energy ranges 
denoted with dotted horizontal lines overlap with some ranges denoted with solid lines.
Upper panel: rms amplitude (with $1\sigma$ error or $1\sigma$ upper limit) versus energy with 
ranges 2.06--3.68 keV (Fig.~\ref{PDS_SAT1}, panel a), 3.68--4.90 keV (Fig.~\ref{PDS_SAT1}, panel b), 4.90--5.71 keV (Fig.~\ref{PDS_SAT1}, panel c), 5.71--6.53 keV (Fig.~\ref{PDS_SAT1}, panel d), 6.53--7.35 keV (Fig.~\ref{PDS_SAT1}, panel e), 7.35--8.98 keV (Fig.~\ref{PDS_SAT1}, panel f), 8.17--10.63 keV (Fig.~\ref{PDS_SAT2}, panel a), 8.98--10.63 keV (Fig.~\ref{PDS_SAT2}, panel b), 8.98--12.28 keV (Fig.~\ref{PDS_SAT2}, panel c), 9.81--13.11 keV (Fig.~\ref{PDS_SAT2}, panel d) and 10.63--20.62 keV (Fig.~\ref{PDS_SAT2}, panel f). This panel shows that the rms amplitude
clearly and gradually decreases at higher energies.
Lower panel: Q-value (corrected for frequency drift with shift-and-add technique for 400 s segments; with $1\sigma$ error) versus energy with
ranges 2.06--3.68 keV (Fig.~\ref{PDS_SAT1}, panel a), 3.68--4.90 keV (Fig.~\ref{PDS_SAT1}, panel b), 4.90--5.71 keV (Fig.~\ref{PDS_SAT1}, panel c), 5.71--6.53 keV (Fig.~\ref{PDS_SAT1}, panel d), 6.53--7.35 keV (Fig.~\ref{PDS_SAT1}, panel e), 7.35--8.98 keV (Fig.~\ref{PDS_SAT1}, panel f), 8.17--10.63 keV (Fig.~\ref{PDS_SAT2}, panel a), 8.98--10.63 keV (Fig.~\ref{PDS_SAT2}, panel b) and 8.98--12.28 keV (Fig.~\ref{PDS_SAT2}, panel c). This panel shows that the 
Q-value does not significantly depend on energy (see \S~\ref{DataAnalysis} and \ref{Results}).
\label{RMS-Qfactor-vs-Energy}}
\end{figure*}

\begin{figure*}
\centering
\includegraphics*[width=0.8\textwidth]{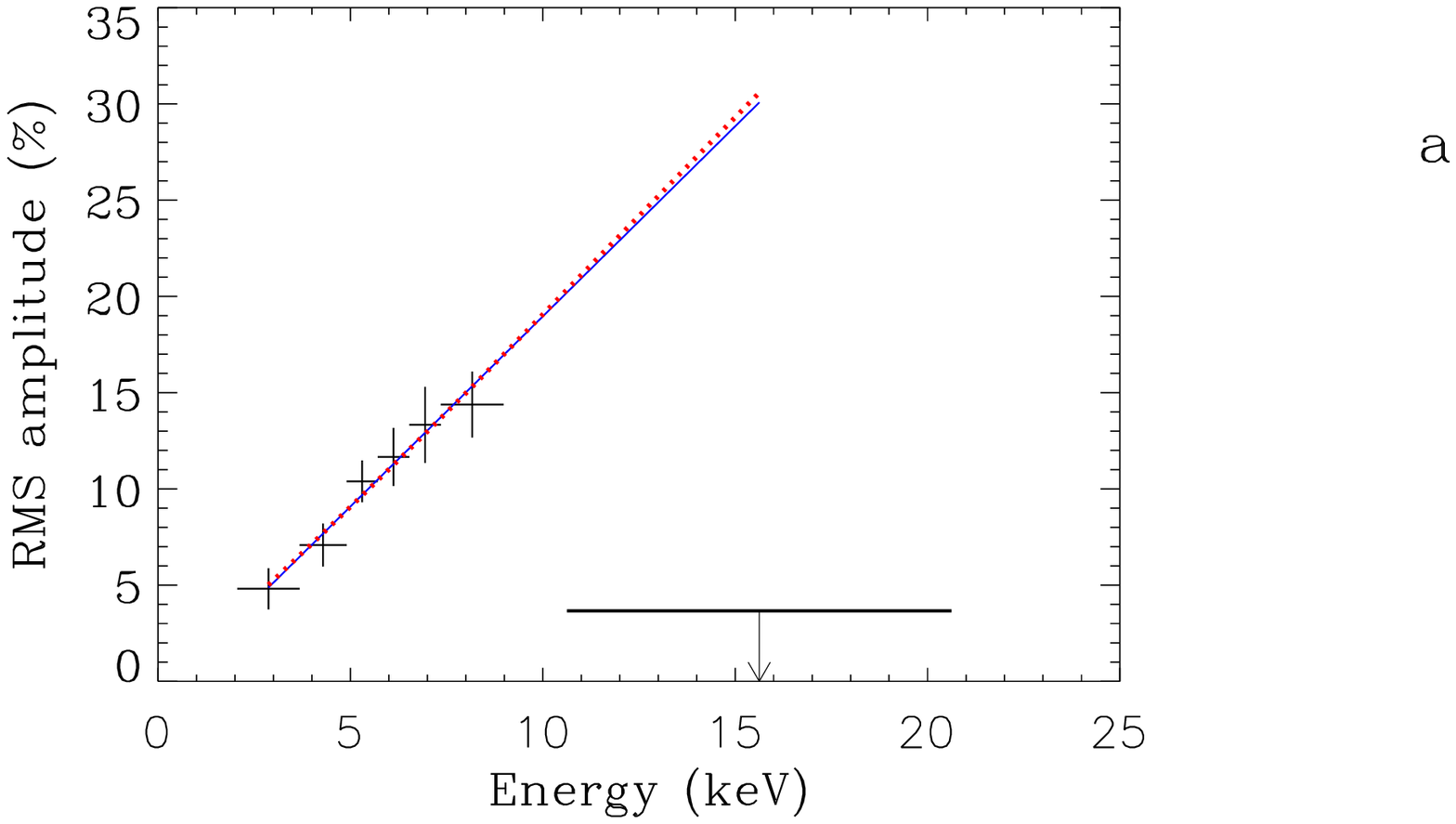}
\caption{Energy dependence of the background corrected fractional rms amplitude
of the lower kHz QPO from 4U 1728--34. The first six data points (with $1\sigma$ errors),
which are same as the first six data points of the upper panel of Fig.~\ref{RMS-Qfactor-vs-Energy},
are fitted with a linear (solid blue line) and a powerlaw model (dotted red line). Both the
best-fit models (extrapolated) are shown. The last data point (for the same energy range as 
that of the last data point of the upper panel of Fig.~\ref{RMS-Qfactor-vs-Energy}) gives the $3\sigma$
upper limit of fractional rms amplitude. This figure shows that the increasing trend 
of amplitude at lower energies cannot explain the last data point.
\label{fit_RMS-Qfactor-vs-Energy_extpl}}
\end{figure*}

In this paper, we report the decrease of fractional rms amplitude of lower
kHz QPOs at higher energies, for the first time to the best of our knowledge.
As the count rate of neutron star LMXBs decreases at higher energies, it is
difficult to reliably measure the fractional rms amplitude
with {\it RXTE} PCA (the only instrument capable of detecting kHz QPOs)
due to the lack of sufficient S/N. Sometimes because of this, the previous studies,
to the best of our knowledge, have concluded that the fractional rms amplitude
either goes on increasing with energies, or first increases and then saturates.
In order to improve the S/N, we combine tens of lower kHz QPOs from
4U 1728--34, and also use the standard shift-and-add technique (\S~\ref{DataAnalysis}).
The net count rate of our highest energy range ($10.63-20.62$ keV)
is also larger than that in some other energy ranges.
Since the $10.63-20.62$ keV range does not have any clear kHz QPO peak,
the decrease of fractional rms amplitude at higher energies should be
real, and not a result of low S/N (see Figs.~\ref{PDS_SAT1} and \ref{PDS_SAT2}).
Note that the small uncertainties in the channel-energy conversion cannot
make the lower kHz QPO disappear at higher energies, and hence cannot
produce the decrease. Moreover, the gradual and systematic change of
fractional rms amplitude at higher energies for overlapping energy ranges
(Figs.~\ref{PDS_SAT1}, \ref{PDS_SAT2}, \ref{RMS-Qfactor-vs-Energy}) strongly
supports the decrease of this amplitude of lower kHz QPOs.

\begin{figure*}
\centering
\begin{tabular}{c}
\hspace{-1.0cm}
\includegraphics*[width=0.8\textwidth]{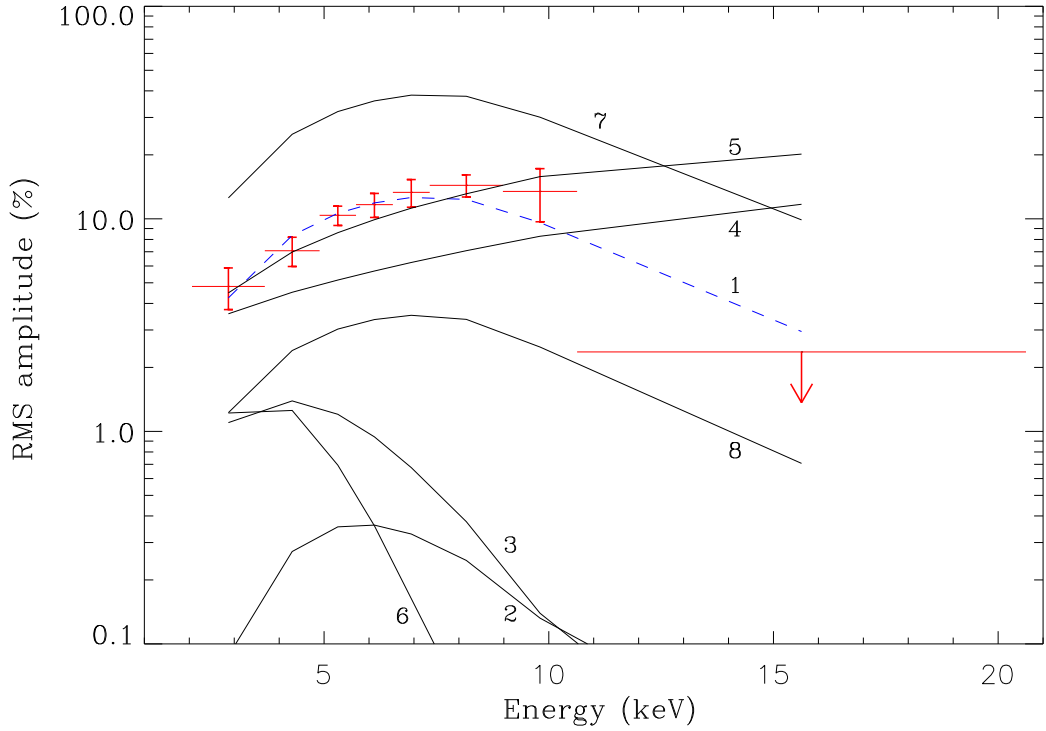}
\end{tabular}
\caption{The lower kHz QPO fractional (in \%) rms amplitude vs. energy. The
red points are the same non-overlapping data points of the upper panel 
of Fig.~\ref{RMS-Qfactor-vs-Energy} (\S~\ref{DataAnalysis} and \ref{Results}).
The curves are models for sinusoidal fluctuation of
the blackbody temperature (see \S~\ref{Models}). The parameter ($N_{\rm PL}$,
$\alpha_{\rm PL}$, $N_{\rm BB}$, $T_{\rm BB}$ and $A$) values for various
model curves are as follows in the same sequence (for units see \S~\ref{Models}).
1: [0.5, 3.5, 50.0, 0.89, 0.034]; 2: [0.5, 3.5, 0.5, 0.88, 0.034];
3: [0.5, 1.5, 50.0, 0.88, 0.034]; 4: [0.5, 3.5, 50.0, 3.0, 0.034];
5: [0.5, 3.5, 50.0, 1.5, 0.034]; 6: [0.5, 3.5, 50.0, 0.5, 0.034]; 7: [0.5, 3.5, 50.0, 0.88, 0.1];
8: [0.5, 3.5, 50.0, 0.88, 0.01]. The blue dashed curve (curve 1) 
appears to be the closest to the data among our computed model curves for 
blackbody temperature fluctuation (see \S~\ref{Models}). This figure shows that
blackbody temperature fluctuation could reproduce the observed energy dependence of
the lower kHz QPO fractional rms amplitude (\S~\ref{Models}).
\label{khzqpoBBpl_BBtemp_1.ps}}
\end{figure*}

\begin{figure*}
\centering
\begin{tabular}{c}
\hspace{-1.0cm}
\includegraphics*[width=0.8\textwidth]{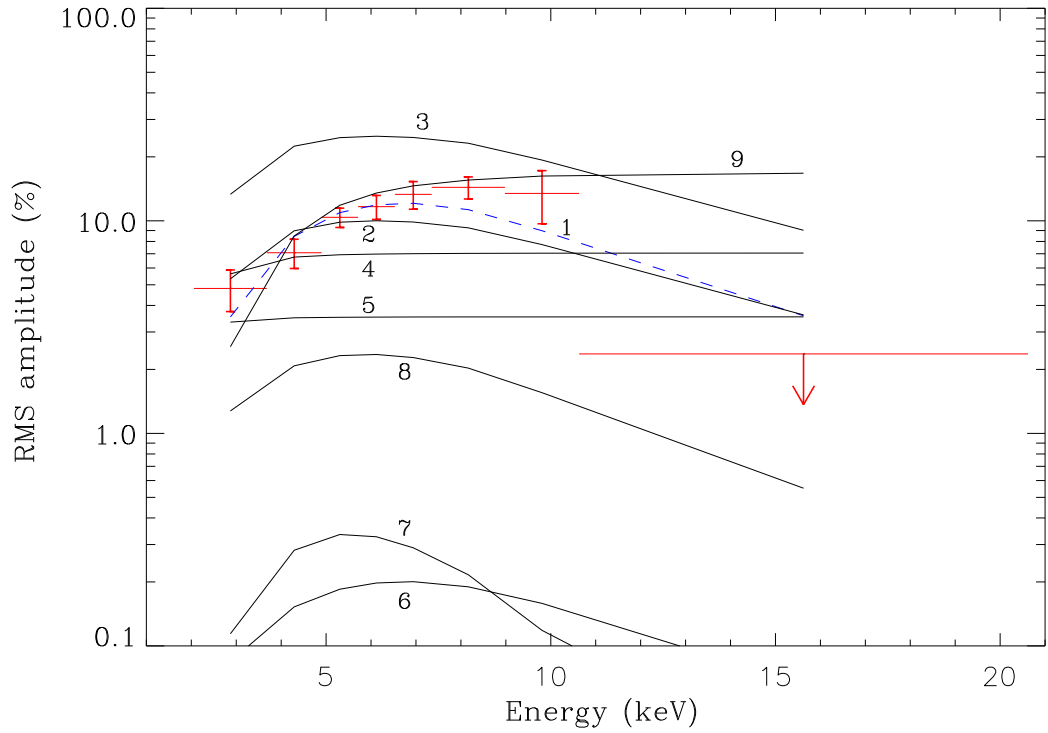}
\end{tabular}
\caption{Similar to Fig.~\ref{khzqpoBBpl_BBtemp_1.ps}, but the curves are models
for sinusoidal fluctuation of the blackbody normalization (see \S~\ref{Models}).
The parameter ($N_{\rm PL}$,
$\alpha_{\rm PL}$, $N_{\rm BB}$, $T_{\rm BB}$ and $A$) values for various
model curves are as follows in the same sequence (for units see \S~\ref{Models}).
1: [3.69, 2.79, 10.37, 1.42, 0.9]; 2: [0.5, 3.5, 13.8, 1.11, 0.2]; 3: [0.5, 3.5, 13.8, 1.11, 0.5];
4: [0.5, 3.5, 12.4, 3.0, 0.1]; 5: [0.5, 3.5, 50.0, 3.0, 0.05]; 6: [10.0, 1.5, 10.0,
2.0, 0.1]; 7: [10.0, 3.5, 10.0, 1.0, 0.1]; 8: [0.5, 2.0, 10.0, 1.5, 0.1];
9: [50.0, 3.5, 50.0, 3.0, 0.25].
The blue dashed curve (curve 1) appears to be
the closest to the data among our computed model curves for
blackbody normalization fluctuation (see \S~\ref{Models}). This figure shows that
blackbody normalization fluctuation could reproduce the observed energy dependence of
the lower kHz QPO fractional rms amplitude, although for a high value of $A$ (\S~\ref{Models}).
\label{khzqpoBBpl_BBNorm_1.ps}}
\end{figure*}

\begin{figure*}
\centering
\begin{tabular}{c}
\hspace{-1.0cm}
\includegraphics*[width=0.8\textwidth]{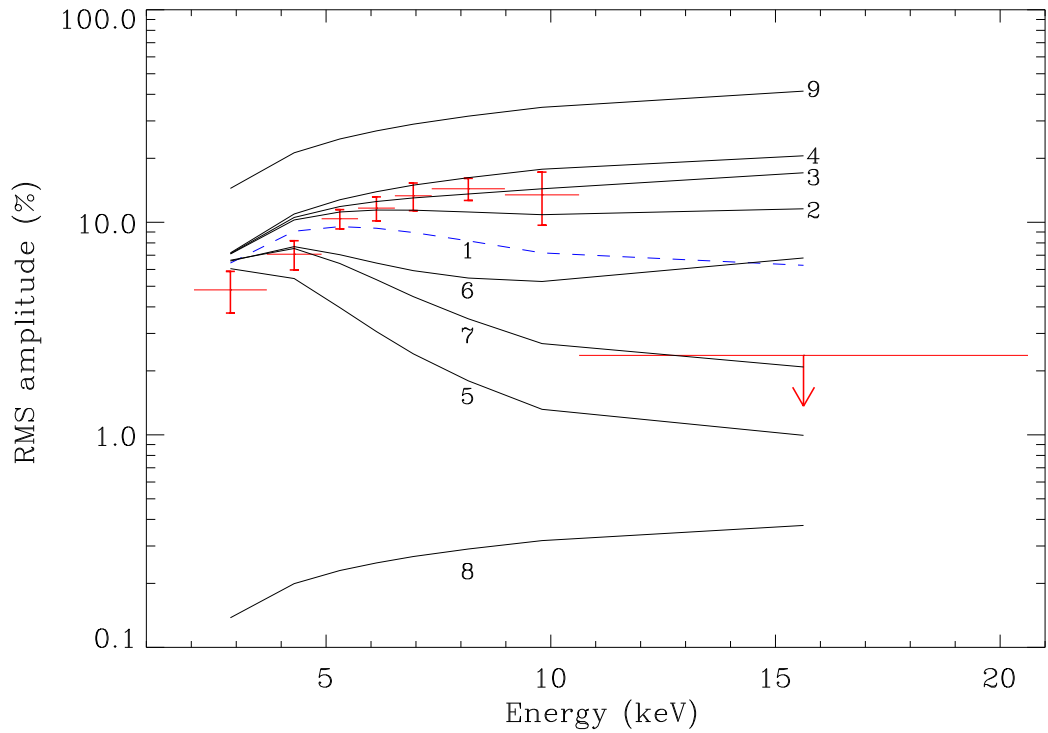}
\end{tabular}
\caption{Similar to Fig.~\ref{khzqpoBBpl_BBtemp_1.ps}, but the curves are models
for sinusoidal fluctuation of the powerlaw index (see \S~\ref{Models}).
The parameter ($N_{\rm PL}$,
$\alpha_{\rm PL}$, $N_{\rm BB}$, $T_{\rm BB}$ and $A$) values for various
model curves are as follows in the same sequence (for units see \S~\ref{Models}).
1: [18.55, 3.5, 2.07, 3.0, 0.027]; 2: [45.0, 3.5, 5.0, 2.5, 0.03]; 3: [45.0, 3.5, 5.0, 2.0, 0.03];
4: [45.0, 3.5, 5.0, 1.0, 0.03]; 5: [45.0, 3.5, 50.0, 3.0, 0.03]; 6: [45.0, 3.5, 50.0, 2.0, 0.03];
7: [10.0, 3.5, 5.0, 3.0, 0.03]; 8: [10.0, 2.0, 10.0, 1.5, 0.001]; 9: [10.0, 2.0, 10.0, 1.5, 0.1].
The blue dashed curve (curve 1) appears to be
the closest to the data among our computed model curves for
powerlaw index fluctuation (see \S~\ref{Models}). This figure shows that powerlaw index 
fluctuation plausibly could not reproduce the observed energy 
dependence of the lower kHz QPO fractional rms amplitude (see \S~\ref{Models}).
\label{khzqpoBBpl_plindex_1.ps}}
\end{figure*}

\begin{figure*}
\centering
\begin{tabular}{c}
\hspace{-1.0cm}
\includegraphics*[width=0.8\textwidth]{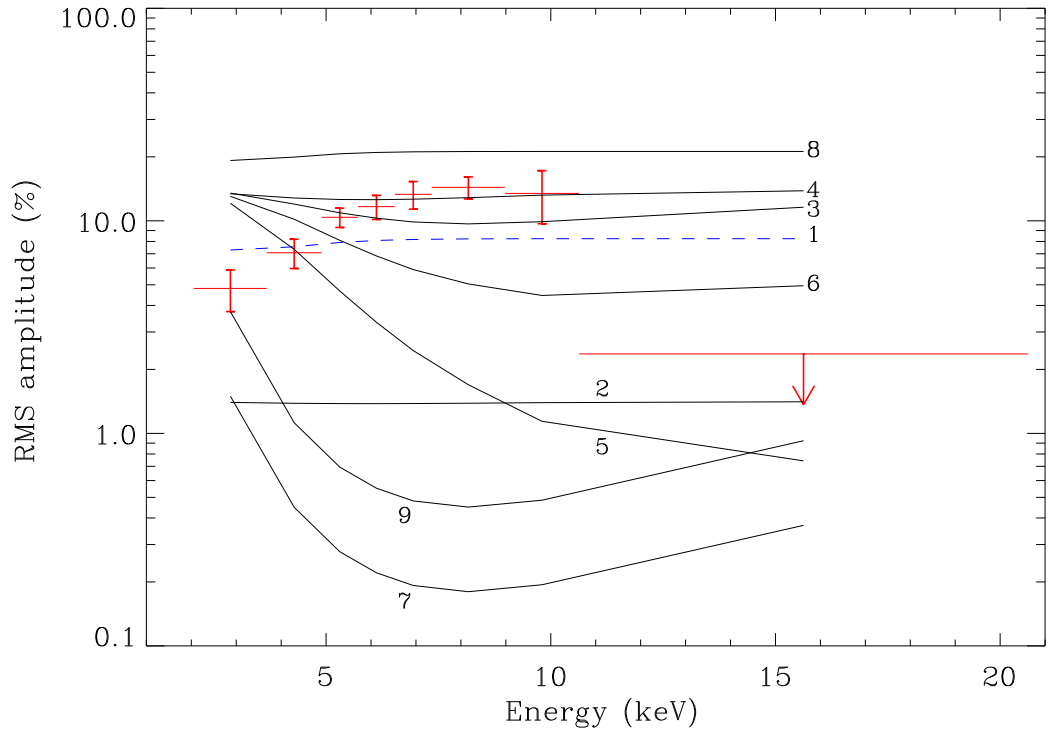}
\end{tabular}
\caption{Similar to Fig.~\ref{khzqpoBBpl_BBtemp_1.ps}, but the curves are models
for sinusoidal fluctuation of the powerlaw normalization (see \S~\ref{Models}).
The parameter ($N_{\rm PL}$,
$\alpha_{\rm PL}$, $N_{\rm BB}$, $T_{\rm BB}$ and $A$) values for various
model curves are as follows in the same sequence (for units see \S~\ref{Models}).
1: [0.5, 3.5, 50.0, 0.52, 0.117]; 2: [10.0, 2.0, 10.0, 1.5, 0.02]; 3: [10.0, 3.5, 10.0, 1.5, 0.2];
4: [10.0, 2.0, 50.0, 1.5, 0.2]; 5: [50.0, 3.5, 50.0, 3.0, 0.2]; 6: [0.5, 3.5, 0.5, 2.0, 0.2];
7: [0.5, 3.5, 50.0, 1.5, 0.12]; 8: [0.5, 3.5, 50.0, 0.5, 0.3]; 9: [0.5, 3.5, 50.0, 1.5, 0.3].
The blue dashed curve (curve 1) appears to be
the closest to the data among our computed model curves for
powerlaw normalization fluctuation (see \S~\ref{Models}). This figure shows that powerlaw 
normalization fluctuation plausibly could not reproduce the observed energy 
dependence of the lower kHz QPO fractional rms amplitude (see \S~\ref{Models}).
\label{khzqpoBBpl_plNorm_1.ps}}
\end{figure*}

The energy dependence of fractional rms amplitude can be very useful to
probe the physical origin of kHz QPOs (see \S~\ref{Introduction}), and
hence our finding is important. Therefore, we use a two-component (blackbody+powerlaw) 
spectral model to gain insights (see \S~\ref{Models} for details).
We explore the shape and fractional rms amplitude values of many model curves,
examples of which have been shown in Figs.~\ref{khzqpoBBpl_BBtemp_1.ps},
\ref{khzqpoBBpl_BBNorm_1.ps}, \ref{khzqpoBBpl_plindex_1.ps} and 
\ref{khzqpoBBpl_plNorm_1.ps}. These curves suggest
that (1) the reported energy behavior of lower kHz QPO fractional rms amplitude
can be explained with simple models; and (2) it is likely that the fluctuation in a
blackbody-like component primarily causes the lower kHz QPOs.
However, we consider the fluctuation of only one model parameter at a time
(\S~\ref{Models}), and hence we cannot study the
effects of a simultaneous fluctuation of the two powerlaw parameters.
We also note that the non-dependence of Q-value on energy (\S~\ref{DataAnalysis})
may imply that the lifetime of the oscillations is independent of photon energies.
This may suggest that the lower kHz QPOs in various energy ranges originate from the
same location, and hence from the same spectral component. But we cannot constrain
the location of either the blackbody or the powerlaw component from our study.
The blackbody could originate either from the boundary layer or the disk
\citep{Taranaetal2011}. Similarly, the corona, which plausibly gives rise to the powerlaw,
could be centrally located near the neutron star or a cover on the accretion disk.

With our new finding, an important question to ask would be whether the fractional 
rms amplitude of lower kHz QPOs decreases with energy (at higher energies) 
for other neutron star LMXBs
as well. \citet{Gilfanovetal2003} reported that 4U 1608--52 did not show any decrease
in fractional rms amplitudes at energies as high as 20 keV,
although at the high energy end the error in fractional rms amplitude was quite high.
A larger (than PCA) area at higher energies may be required (1) to detect a plausible
decrease which happens at higher energies than that for 4U 1728--34; (2) and to
reduce the errors in fractional rms amplitudes, while increasing the number of
data points, in order to perform a better modeling. The upcoming space missions
(e.g., {\it ASTROSAT}) having sufficient time resolution and much higher
(than PCA) area in $20-50$ keV, will be useful for this purpose.

\acknowledgments

We thank Manoneeta Chakraborty and H. M. Antia for technical discussions. A.M. acknowledges Ranjeev Misra for a discussion on the models and thanks M. Coleman Miller for useful general discussions on kHz QPOs. We also thank an anonymous referee for very constructive comments, which significantly improved the paper.

{}

\clearpage

\clearpage

\clearpage

\clearpage

\clearpage

\clearpage

\clearpage

\clearpage


\begin{thebibliography}{}

\bibitem[Abramowicz and Klu\'zniak (2001)]{AbramowiczKluzniak2001} Abramowicz, M. A.,
\& Klu\'zniak, W. 2001, A\&A, 374, L19

\bibitem[Alpar and Psaltis (2008)]{AlparandPsaltis2008} Alpar, M. A., \& Psaltis, D. 2008, MNRAS, 391, 1472


\bibitem[Barret (2001)]{Barret2001} Barret, D. 2001, Advances in Space Research, 28, 307


\bibitem[Barret et al. (2005a)]{Barretetal2005a} Barret, D., Kluźniak, W., Olive, J.-F., 
Paltani, S. \& Skinner, G. K. 2005a, MNRAS, 357, 1288

\bibitem[Barret et al. (2005b)]{Barretetal2005b} Barret, D., Olive, J.-F., \&
Miller, M. C. 2005b, MNRAS, 361, 855

\bibitem[Barret et al. (2006)]{Barretetal2006} Barret, D., Olive, J.-F., \&
Miller, M. C. 2006, MNRAS, 370, 1140

\bibitem[Bhattacharyya (2010)]{Bhattacharyya2010} Bhattacharyya, S. 2010, Advances in Space Research, 45, 949



\bibitem[Cabanac et al. (2010)]{Cabanacetal2010} Cabanac, C., Henri, G., Petrucci, P. -O., Malzac, J., Ferreira, J., \& Belloni, T. M. 2010, 404, 738

\bibitem[Christian and Swank (1997)]{ChristianSwank1997} Christian, D. J., \& Swank, J. H. 1997, ApJ, 109, 177

\bibitem[Church and Baluci\'nska-Church (2001)]{ChurchBalucinskaChurch2001} Church, M. J., \& Baluci\'nska-Church, M. 2001, A\&A, 369, 915

\bibitem[Di Salvo et al. (2001)]{Disalvoetal2001} Di Salvo, T., M\'endez, M., van der Klis, M., 
Ford, E., \& Robba, N. R. 2001, ApJ, 546, 1107



\bibitem[Ford and van der Klis (1998)]{FordvanderKlis1998} Ford, E. C., \& van der Klis, M. 1998, ApJ, 506, L39

\bibitem[Gilfanov et al. (2003)]{Gilfanovetal2003} Gilfanov, M., Revnivtsev, M., \& Molkov, S. 2003, A\&A, 410, 217


\bibitem[Gierli{\'n}ski \& Zdziarski(2005)]{2005MNRAS.363.1349G} Gierli{\'n}ski, M., \& Zdziarski, A.~A.\ 2005, \mnras, 363, 1349 


\bibitem[Kato (2011)]{Kato2011} Kato, S. 2011, Publ. Astron. Soc. Japan 63, 861


\bibitem[Kluzniak \& Abramowicz(2001)]{2001AcPPB..32.3605K} Kluzniak, W., \& Abramowicz, M.~A.\ 2001, Acta Physica Polonica B, 32, 3605 

\bibitem[Lamb and Miller (2003)]{LambMiller2003} Lamb, F. K., \& Miller, M. C. 2003, submitted (astro-ph/0308179)

\bibitem[Leahy et al. (1983)]{Leahyetal1983} Leahy, D. A., Darbro, W., Elsner, R. F.,
Weisskopf, M. C., Sutherland, P. G., Kahn, S., \& Grindlay, J. E., 1983, ApJ, 266, 160

\bibitem[Lee et al. (2004)]{Leeetal2004} Lee, W. H., Abramowicz, M. A., \& Klu\'zniak, W. 2004, ApJ, 603, L93

\bibitem[Lee and Miller (1998)]{LeeMiller1998} Lee, H. C., \& Miller, G. S. 1998, MNRAS, 299, 479

\bibitem[Lin et al. (2011)]{Linetal2011} Lin, Y. F., Boutelier, M., Barret, D., \& Zhang, S. N. 2011, ApJ, 726, 74

\bibitem[Lin et al. (2007)]{Linetal2007} Lin, D., Remillard, R. A., \& Homan, J. 2007, ApJ, 667, 1073

\bibitem[Lin et al. (2010)]{Linetal2010} Lin, D., Remillard, R. A., \& Homan, J. 2010, ApJ, 719, 1350

\bibitem[Maccarone and Coppi (2003)]{MaccaroneCoppi2003} Maccarone, T. J., \& Coppi, P. S. 2003, A\&A, 399, 1151

\bibitem[Maitra and Bailyn (2004)]{MaitraBailyn2004} Maitra, D., \& Bailyn, C. D. 2004, ApJ, 608, 444


\bibitem[M\'endez (2006)]{Mendez2006} M\'endez, M. 2006, MNRAS, 371, 1925

\bibitem[M\'endez and Belloni (2007)]{MendezBelloni2007} M\'endez, M., \& Belloni, T. 2007, MNRAS, 381, 790

\bibitem[M\'endez and van der Klis(1999)]{MendezvanderKlis1999} M\'endez, M., \& van der Klis, M. 1999, ApJ, 517, L51

\bibitem[M\'endez et al. (2001)]{Mendezetal2001} M\'endez, M., van der Klis, M., \& Ford, E. C. 2001, ApJ, 561, 1016

\bibitem[M\'endez et al. (1998)]{Mendezetal1998} M\'endez, M., van der Klis, M., van Paradijs, J., Lewin, W. H. G., Vaughan, B. A., Kuulkers, E., Zhang, W., Lamb, F. K. \& Psaltis, D. 1998, ApJ, 494, L65

\bibitem[Miller et al. (1998)]{Milleretal1998} Miller, M. C., Lamb, F. K., \& Psaltis, D. 1998, ApJ, 508, 791

\bibitem[Mitsuda et al. (1989)]{Mitsudaetal1989} Mitsuda, K., Inoue, H., Nakamura, N., \& Tanaka, Y. 1989, PASJ, 41, 97

\bibitem[Mitsuda and Dotani (1989)]{MitsudaandDotani1989} Mitsuda, K., \& Dotani, T. 1989, PASJ, 41, 557

\bibitem[Mukherjee and Bhattacharyya (2011a)]{MukherjeeBhattacharyya2011a} Mukherjee, A., \& Bhattacharyya, S. 2011, MNRAS, 411, 2717

\bibitem[Mukherjee and Bhattacharyya (2011b)]{MukherjeeBhattacharyya2011b} Mukherjee, A., \& Bhattacharyya, S. 2011, ApJ, 730, L32

\bibitem[Mukhopadhyay (2009)]{Mukhopadhyay2009} Mukhopadhyay, B. 2009, ApJ, 694, 387

\bibitem[Muno et al. (2002)]{Munoetal2002} Muno, M. P., \"Ozel, F., \& Chakrabarty, D. 2002, ApJ, 581, 550

\bibitem[Olive et al. (2003)]{Oliveetal2003} Olive, J. F., Barret, D., \& Gierl\'nski, M. 2003, 583, 416

\bibitem[P\'etri (2011)]{Petri2011} P\'etri, J. 2011, Astrophysics and Space Science, 331, 555


\bibitem[Psaltis (2008)]{Psaltis2008} Psaltis, D. 2008, Living Reviews in Relativity, 11, 9

\bibitem[Romanova and Kulkarni (2009)]{Romanovakulkarni2009} Romanova, M. M., \& Kulkarni, A. K. 2009, MNRAS,
398, 1105

\bibitem[Seifina and Titarchuk (2011)]{SeifinaTitarchuk2011} Seifina, E., \& Titarchuk, L. 2011, ApJ, 738, 128

\bibitem[Stella and Vietri (1998)]{StellaVietri1998} Stella, L., \& Vietri, M. 1998, ApJ, 492, L59

\bibitem[Stella and Vietri (1999)]{StellaVietri1999} Stella, L., \& Vietri, M. 1999, Physical Review Letters, 82, 17

\bibitem[Strohmayer et al. (1996)]{Strohmayeretal1996} Strohmayer, T. E., Zhang, W., Swank, J. H., Smale, A., Titarchuk, L., Day, C., \& Lee, U. 1996, ApJ, 469, L9

\bibitem[Stuchl\'ik et al. (2011)]{Stuchliketal2011} Stuchl\'ik, Z., Kotrlov\'a, A., \& T\..or\..ok, G. 2011, A\&A, 525, 82

\bibitem[Tarana et al. (2011)]{Taranaetal2011} Tarana, A., Belloni, T., Bazzano, A., Méndez, M., \& Ubertini, P. 2011, MNRAS, 416, 873

\bibitem[T\"or\"ok (2009)]{Torok2009} T\"or\"ok, G. 2009, A\&A, 497, 661

\bibitem[van der Klis (1989)]{vanderKlis1989} van der Klis, M. 1989, Eds. H. \"Ogelman and E.P.J. van den Heuvel, (Kluwer Academic Publishers: Boston), 27


\bibitem[van der Klis(2006)]{vanderKlis2006} van der Klis, M.\ 2006, Compact stellar X-ray sources, 39 

\bibitem[van der Klis et al. (1996)]{vanderKlisetal1996} van der Klis, M., Swank, J. H., \& Zhang, W. 1996, ApJ, 469, L1

\bibitem[van der Klis et al. (1996)]{vanderKlisetal1996} van der Klis, M., Swank, J. H., Zhang, W., Jahoda, K., Morgan, E. H., Lewin, W. H. G., Vaughan, B., \& van Paradijs, J. 1996, ApJ, 469, L1


\bibitem[Wang et al. (2011)]{Wangetal2011} Wang, J., Zhang, C. M., Zhao, Y. H., Lin, Y. F., Yin, H. X., \& Song, L. M. 2011, A\&A, 528, 126

\bibitem[White et al. (1988)]{Whiteetal1988} White, N. E., Stella, L., \& Parmar, A. N. 1988, ApJ, 324, 363

\bibitem[Wijnands (2001)]{Wijnands2001} Wijnands, R. 2001, Advances Space Res., 28, 469

\bibitem[Wijnands et al. (2003)]{Wijnandsetal2003} Wijnands, R., van der Klis, M., Homan, J., Chakrabarty, D., Markwardt, C. B., \& Morgan, E. H. 2003, Nature, 424, 44

\bibitem[Zdziarski et al.(2005)]{2005MNRAS.360..825Z} Zdziarski, A.~A., Gierli{\'n}ski, M., Rao, A.~R., Vadawale, S.~V., \& Miko{\l}ajewska, J.\ 2005, MNRAS, 360, 825 

\bibitem[Zhang (2004)]{Zhang2004} Zhang, C. 2004, A\&A, 423, 401

\bibitem[Zhang et al. (2006)]{Zhangetal2006} Zhang, C. M., Yin, H. X., Zhao, Y. H., Zhang, F., \& Song, L. M. 2006, MNRAS, 366, 1373

\end{thebibliography}
\end{document}